\documentclass[12pt]{article}

\usepackage{amssymb}
\usepackage{amsmath}
\usepackage{amscd}
\usepackage{latexsym}
\usepackage{graphicx}
\usepackage{url}

\usepackage{cite}

\newcommand{\be}{\begin{equation}}
\newcommand{\ee}{\end{equation}}

\newcommand{\dlt}{\delta}

\newcommand{\al}{\alpha}
\newcommand{\ra}{\rightarrow}

\newcommand{\rgl}{\rangle}
\newcommand{\lgl}{\langle}

\begin{document}

\begin{center}

{\Large{\bf Preference reversal in quantum decision theory} \\ [5mm]

V.I. Yukalov$^{1,2,*}$ and D. Sornette$^{1,3}$} \\ [3mm]

{\it
$^1$Department of Management, Technology and Economics, \\
ETH Z\"urich (Swiss Federal Institute of Technology) \\
Scheuchzerstrasse 7,  Z\"urich CH-8032, Switzerland \\ [3mm]

$^2$Bogolubov Laboratory of Theoretical Physics, \\
Joint Institute for Nuclear Research, Dubna 141980, Russia \\ [3mm]

$^3$Swiss Finance Institute, c/o University of Geneva, \\
40 blvd. Du Pont d'Arve, CH 1211 Geneva 4, Switzerland}

\end{center}

\vskip 3cm

\begin{abstract}

We consider the psychological effect of preference reversal and show
that it finds a natural explanation in the frame of quantum decision
theory. When people choose between lotteries with non-negative 
payoffs, they prefer a more certain lottery because of uncertainty 
aversion. But when people evaluate lottery prices, e.g. for selling 
to others the right to play them, they do this more rationally, being 
less subject to behavioral biases. This difference can be explained by
the presence of the attraction factors entering the expression of quantum
probabilities. Only the existence of attraction factors can explain why,
considering two lotteries with close utility factors, a decision maker 
prefers one of them when choosing, but evaluates higher the other one 
when pricing. We derive a general quantitative criterion for the 
preference reversal to occur that relates the utilities of the two 
lotteries to the attraction factors under choosing versus pricing and 
test successfully its application on experiments by Tversky et al.
We also show that the planning paradox can be treated as a kind of 
preference reversal.

\end{abstract}

\vskip 2cm

{\parindent=0pt
{\bf Keywords}: preference reversal, decision theory, uncertainty,
behavioral quantum probability, planning paradox

\vskip 1cm

{\bf *Correspondence}: \\
V.I. Yukalov \\
{\it
Department of Management, Technology and Economics \\
ETH Z\"urich (Swiss Federal Institute of Technology) \\
Scheuchzerstrasse 7,  Z\"urich CH-8032, Switzerland }
\vskip 1mm
{\bf e-mail}: syukalov@ethz.ch

}

\newpage

\section{Introduction}

For many decades, psychologists and economists have been intrigued by 
a seemingly anomalous effect termed {\it preference reversal}. The 
simplest example illustrating this effect is as follows. First, subjects 
are asked to choose between two lotteries, say $L_1$ and $L_2$, such 
that $L_1$ has a high chance to win a relatively modest prize, while 
$L_2$ offers a lower chance of winning, but an essentially larger prize. 
The majority of subjects choose the more certain win of lottery $L_1$, 
despite the fact that lottery $L_2$ can enjoy a larger expected utility. 
Then subjects are asked to price each of the lotteries, as if they would 
own them and wish to sell the right to play them. Surprisingly, the 
majority of subjects price higher the less certain lottery $L_2$ in 
apparent contradiction with their previous choice. This example embodies 
the essence of the preference reversal effect.

Among the first scientists emphasizing the existence of this effect were
Lindman (1971) and Lichtenstein and Slovic (1971, 1973). Their studies were
followed by several authors demonstrating the occurrence of this effect
in psychology and economics (Grether and Plott, 1979;
Loomes and Sugden, 1983; Holt, 1986; Goldstein and Einborn, 1987;
Karni and Safra, 1987; Segal, 1988; Tversky et al., 1988;
Schkade and Johnson, 1989). Many other citations can be found in the review
articles (Slovic and Lichtenstein, 1983; Tversky and Thaler, 1990,
Tversky et al., 1990). The experimental studies have established the clear
validity and robustness of the preference reversal phenomenon.

The preference reversal effect looks surprising because, according to the
common understanding of utility, the choice among the given lotteries
should be based on the objective values of the latter, thus, being procedure
invariant. Since the lottery values are not changed, why then is the
preference reversed?

It has been proved by Tversky and Thaler (1990) and Tversky et al. (1990)
that it is the {\it breaking of procedure invariance} that is responsible
for the preference reversal phenomenon. It turns out that subjects weight
more heavily payoffs in pricing than in choice, so that the preference
reversal is a purely psychological effect.

The origin of the preference reversal has been recently explained from the
point of view of neurology by Kim et al. (2012). It has been experimentally
shown that there exists correlation between visual fixation and preferences.
Visual fixations both reflect and influence preferences. From one side, these
fixations reflect which objects seem to be more important for the subject.
And, from the other side, such fixations modulate the neural correlates of
preferences, with activity in ventromedial prefrontal cortex and ventral
striatum, reflecting the value of the fixated item compared to the value
of the item not fixated. Kim et al. studied the process of decision making
under risk and measured eye movements while people chose between gambles
or bid in pricing gambles. Consistently with the previous work, they found
that, for two gambles matched in expected value, people systematically chose
the higher probability option, but requested a higher ask price for the 
option that offered the greater amount to win, thus demonstrating preference
reversal. 

This effect was accompanied by a shift in fixation of the two attributes, 
with people fixating more on probabilities during choices  and more on 
amounts during selling. In this way, there exists 
{\it probability-versus-amount dichotomy}: When choosing, one pays more 
attention to probabilities while, when selling, one better appreciates 
amounts.

Understanding the cause of the preference reversal is the first 
necessary step. The next step should be the description of this effect 
by a mathematical model. Previous suggested models were not successful, 
as was analyzed by Tversky and Thaler (1990) and Tversky et al. (1990). 
In the present paper, we show that the effect of preference reversal 
finds a simple and natural explanation in the frame of the Quantum 
Decision Theory developed by the authors 
(Yukalov and Sornette, 2008, 2009a, 2009b, 2010,2011, 2013, 2014a, 2014b, 2015).

\section{Basics of quantum decision theory}

There exists several approaches applying quantum notions to psychological
sciences, as can be inferred from the books (Khrennikov 2010;
Busemeyer and Bruza 2012; Haven and Khrennikov 2013; Bagarello 2013) and the
review articles (Yukalov and Sornette 2009b; Sornette 2014; 
Busemeyer et al. 2014; Ashtiani and Azgomi 2015), where numerous citations 
to the previous literature can be found. Quantum Decision Theory (QDT) 
principally differs from all those approaches in two aspects. First, QDT is 
based on a self-consistent mathematical foundation that is common for both 
quantum measurement theory and quantum decision theory. Starting from the 
von Neumann (1955) theory of quantum measurements, we have generalized it 
to the case of uncertain or inconclusive events, making it possible to 
characterize uncertain measurements and uncertain prospects. Second, the main 
formulas of QDT are derived from general principles, giving the possibility 
of quantitative predictions, without fitting parameters. This is in contrast 
with the usual way of constructing particular models for describing some 
concrete experiments, with fitting the model parameters from empirical data.

We shall not repeat here the mathematical foundation of QDT that has been
thoroughly expounded in our previous papers, but we will just briefly recall
the resulting formulas that are necessary for describing the 
preference reversal effect.

Let us consider a composite event, called prospect,
\be
\label{1}
 \pi_n = A_n \bigotimes B \;  .
\ee
Here $A_n$ is an operationally testable event, represented in a Hilbert space
by an eigenstate $\vert n \rangle$. While $B = \{ B_\alpha,\; b_\alpha \}$ is
an inconclusive event that is a set of possible events $B_\alpha$, represented
in a Hilbert space by eigenstates $\vert \alpha \rangle$, and equipped with
random amplitudes $b_\alpha$, so that the inconclusive event is represented 
by a state $\vert B \rangle = \sum_\alpha b_\alpha \vert \alpha \rangle$.

The prospect operator is $\hat{P}(\pi_n)=\vert n B \rangle\langle n B \vert$,
such that the prospect probability is given by the quantum formula
\be
\label{2}
 p(\pi_n) = {\rm Tr} \hat\rho \hat P(\pi_n) \;  ,
\ee
where $\hat{\rho}$ is a strategic state of a decision maker. By
construction, the prospect probability enjoys the properties of a
probability measure:
\be
\label{3}
 \sum_n p(\pi_n) = 1 \; , \qquad 0 \leq p(\pi_n) \leq 1 \; .
\ee

It is easy to show that the prospect probability takes the form
\be
\label{4}
 p(\pi_n) = f(\pi_n) + q(\pi_n) \;  ,
\ee
where the first term is called {\it utility factor}, characterizing the 
utility of the prospect, while the second term is {\it attraction factor} 
representing behavioral biases.

The intuitive explanation of the above probability expression (\ref{4}) is 
straightforward: The definition of a quantum probability (\ref{2}) for a 
composite event can be separated into a term containing diagonal matrix 
elements and a term including off-diagonal elements. The diagonal elements
compose the term $f(\pi_n)$, while the off-diagonal elements define the term
$q(\pi_n)$. The occurrence of an off-diagonal term is a typical feature
of quantum theory, where this quantity is called {\it interference term}
or {\it coherence term}. The existence of such an interference term 
constitutes the principal difference of the quantum approach from the 
classical consideration, where there are no interference terms. It is the 
appearance of interference terms that makes the structure of quantum 
expressions richer then the related classical ones and that allows one to 
explain those psychological phenomena that, otherwise, are inexplicable 
in classical decision making. Sometimes, the quantum approach even yields 
conclusions that are impossible in classical decision making, as, for 
instance, the possibility to agree on disagree (Khrennikov and Basieva, 2014). 
Below we show that this interference term, composing the attraction factor, 
is essential in explaining the existence of the preference reversal effect
that cannot be described in classical decision theory.    

The prospect probability satisfies the quantum-classical correspondence
principle.
\be
\label{5}
 p(\pi_n) \ra  f(\pi_n) \; , \qquad q(\pi_n) \ra 0 \; .
\ee
This defines the utility factor as a classical-type probability, with the
standard properties
\be
\label{6}
 \sum_n f(\pi_n) = 1 \; , \qquad 0 \leq f(\pi_n) \leq 1 \;  .
\ee
This is equivalent to the normalization condition
$$
 \sum_{n\al} | b_\al |^2
\lgl n \al \; | \; \hat\rho \; | \; n \al \rgl = 1 \; ,
$$
imposing a constraint on the random quantities $b_\al$.  

When considering lotteries, an event $A_n\equiv A(L_n)$ implies the
choice of a lottery $L_n$. Then the inconclusive set $B$ characterizes
the decision maker hesitations between uncertain events $B_\alpha$,
describing uncertainty with respect to the decision maker ability and
with respect to the lottery formulation (Yukalov and Sornette, 2014b, 2015).
The explicit form of the utility factor is given by minimizing the
Kullback-Leibler information functional, which in the simple case of
uncertainty yields
\be
\label{7}
 f(\pi_n) = \frac{U(L_n)}{\sum_n U(L_n) }  \;  ,
\ee
with $U(L_n)$ being the expected utility of a lottery $L_n$. Note that
the minimization of the information functional results in expression
(\ref{7}) that might be familiar to psychologists as a Luce (1959) 
choice rule using utility as response strength.

The attraction factor reflects the effects of quantum coherence and
interference, and in decision theory it represents the behavioral biases
rendering the prospects more or less attractive from the subconscious
point of view of decision maker. By their definition, attraction factors
lie in the interval
\be
\label{8}
 -1 \leq q(\pi_n) \leq 1
\ee
and satisfy the {\it alternation property}
\be
\label{9}
 \sum_n q(\pi_n) = 0 \;  .
\ee
Also, in the case of non-informative priors, the attraction factors for
the considered prospect lattice $\{\pi_n: n= 1,2,\ldots,N\}$ obey the
{\it quarter law}
\be
\label{10}
 \frac{1}{N} \sum_{n=1}^N | q(\pi_n) | = \frac{1}{4} \;  .
\ee
This law makes it admissible to estimate the attraction factors by the
values $\pm 0.25$, thus quantitatively predicting preferences.

The prospect lattice is ordered by the values of prospect probabilities.
A prospect $\pi_i$ is termed preferable to $\pi_j$ if and only if
$$
 p(\pi_i) > p(\pi_j) \qquad ( \pi_i > \pi_j) \; .
$$
At the same time, a prospect $\pi_i$ is more useful than $\pi_j$ when
$f(\pi_i) > f(\pi_j)$. A prospect $\pi_i$ is more attractive than $\pi_j$,
when $q(\pi_i) > q(\pi_j)$. In this way, a prospect can be more useful but
less attractive, as a result being less preferable.

A necessary condition for the existence of a nonzero attraction factor is
that the composite prospect be entangled (Yukalov and Sornette, 2014a, 2015).
Otherwise, there is no need of involving quantum probabilities.

\section{General criterion of preference reversal}

Preference reversal may naturally arise in the frame of quantum decision
theory. In this section, we derive the general criterion for the occurrence
of this effect.

Suppose a decision maker considers a lattice of just two prospects
\be
\label{11}
 \pi_n = A(L_n) \bigotimes B \qquad ( n = 1,2 ) \;  ,
\ee
with the intention of choosing between them. Here $A(L_n)$ implies the action
of choosing a lottery $L_n$. And $B$ is a set incorporating uncertainties
associated with this choice. Let one prefer the prospect $\pi_1$ against
$\pi_2$, which means that
\be
\label{12}
  p(\pi_1) > p(\pi_2) \qquad ( \pi_1 > \pi_2) \;  .
\ee
Taking into account the alternation property, we have
\be
\label{13}
 q(\pi_1) + q(\pi_2) = 0 \;  .
\ee
This tells us that the prospect $\pi_1$ is preferred to $\pi_2$ if and only if
\be
\label{14}
  f(\pi_2) - f(\pi_1) < 2 q(\pi_1) \; .
\ee

Now, assume that the decision maker plans to price the given lotteries, e.g.,
wishing to sell them. The lotteries remain the same as before. However, 
uncertainties in selling are of course different from those when choosing, 
hence, the uncertain set $B^\prime$, associated with selling, is different 
from the set $B$ including uncertainties associated with choosing. Now, the 
decision maker evaluates the two different prospects
\be
\label{15}
 \pi_n = A(L_n) \bigotimes B' \qquad ( n = 3,4 ) \;  ,
\ee
where $L_1 = L_3$ and $L_2 = L_4$.

Preference reversal implies that, contrary to the situation with choosing,
now the decision maker evaluates higher the prospect $\pi_4$ compared to
$\pi_3$, so that
\be
\label{16}
 p(\pi_3) < p(\pi_4) \qquad ( \pi_3 < \pi_4) \;  .
\ee
In view of the alternation property
\be
\label{17}
 q(\pi_3) + q(\pi_4) = 0 \;  ,
\ee
the preference of $\pi_4$ occurs only when
\be
\label{18}
 f(\pi_4) - f(\pi_3) < 2 q(\pi_3) \;  .
\ee

Since the lotteries are the same ($L_1 = L_3$ and $L_2 = L_4$), their 
expected utilities are pairwise equal: $U(L_1) = U(L_3)$ and
$U(L_2) = U(L_4)$. Therefore the utility factors are also pairwise equal
\be
\label{19}
 f(\pi_1) = f(\pi_3) \; , \qquad f(\pi_2) = f(\pi_4) \;  .
\ee
Combining the above conditions, we obtain the {\it preference reversal
criterion}:
\be
\label{20}
  2 q(\pi_3) \; < \; f(\pi_2) - f(\pi_1) \; < \; 2 q(\pi_1) \; .
\ee

Let us stress that in classical decision making, where
$q(\pi_1) = q(\pi_3) \equiv 0$, the inequalities (\ref{20}) cannot hold,
which means that it is impossible to suggest a self-consistent mathematical
explanation of the preference reversal phenomenon in classical terms, which
is in agreement with discussions by Tversky and Thaler (1990) and
Tversky et al. (1990). 

Criterion (\ref{20}) not only explains the preference reversal phenomenon, 
but it also provides a quantitative estimate of how likely it may 
happen, as well as a posteriori confirmation of why it has happened. This 
is because the attraction factors are not just some additional arbitrary
characteristics, but because their signs are prescribed by the risk aversion
notion, while their values are constrained by conditions (\ref{8}) to 
(\ref{10}). Thus, due to risk aversion when facing several choices, the 
more certain lottery is more attractive, hence $q(\pi_1) > q(\pi_2)$, which, 
in view of the alternation property (\ref{9}), implies that $q(\pi_1) > 0$, 
while $q(\pi_2) < 0$. Contrary to this, when pricing, risk aversion is 
absent, hence more attractive is the lottery that can provide the larger 
gain, so that $q(\pi_4) > q(\pi_3)$, which, again taking into account the 
alternation property (\ref{9}), tells us that $q(\pi_3) < 0$ while 
$q(\pi_4) > 0$. Estimating the absolute values of the attraction factors by 
the quantity $0.25$, which follows from the quarter law (\ref{10}), we have 
the criterion
$$
 - \frac{1}{2} \; < \; f(\pi_2) - f(\pi_1) \; < \frac{1}{2} \;  \; .
$$     
Therefore, if the given lotteries are such that their utility factors
satisfy the above inequalities, we may expect that preference reversal
can occur. And, vice versa, if preference reversal has happened, then
the above inequalities must hold. Below we demonstrate that criterion 
(\ref{20}) really provides a necessary and sufficient conditions for
the preference reversal phenomenon.

\section{Confirmation of preference reversal criterion}

To confirm the validity of the preference reversal criterion, let us test 
it with empirical data of decision-making experiments. We shall consider 
pairs of lotteries with the notation of the previous section. The prospects,
related to the choice between the lotteries $L_1$ and $L_2$, are denoted as
$\pi_1$ and $\pi_2$, respectively. The prospects, corresponding to
pricing of these lotteries, will be denoted by $\pi_3$ and $\pi_4$. The
expected utility of a lottery $L = \{x_i,p(x_i)\}$, consisting of payoffs
$x_i$, with their weights $p(x_i)$, will be calculated by the formula
$U(L) = \sum_i x_i p(x_i)$. And the utility factors are given by
expression (\ref{7}).

\vskip 2mm

{\bf Example 1}. Let us start with the example given by Tversky and Thaler
(1990). Consider two lotteries
$$
L_1 = \left \{ 4 , \frac{8}{9} \; | \; 0 , \frac{1}{9} \right \} \; ,
\qquad
L_2 = \left \{ 40 , \frac{1}{9} \; | \; 0 , \frac{8}{9} \right \} \; ,
$$
whose payoffs $4$ and $40$ are given in some monetary units. The type of
units, whether these are Dollars, or Euro, or Francs, is not of importance,
since such units are canceled in definition (\ref{7}) of utility factors.
This is one of the advantage of employing the dimensionless utility factors 
that are invariant with respect to the type of payoff measures. The 
corresponding expected utilities
$$
 U(L_1) = \frac{32}{9} \; , \qquad U(L_2) = \frac{40}{9} \;  ,
$$
result in the utility factors
$$
f(\pi_1) = \frac{4}{9} \; , \qquad f(\pi_2) = \frac{5}{9} \;  ,
$$
which show that the second lottery is more useful.

The experimental probabilities are defined as the fractions of subjects
preferring the related lotteries. According to Tversky and Thaler (1990),
in the case of choice, it was found that 71\% of decision makers preferred 
the more certain lottery $L_1$, so that
$$
 p(\pi_1) =  0.71 ~ > ~ p(\pi_2) =  0.29 \; ,
$$
despite that this lottery is less useful. In view of (\ref{4}), this 
corresponds to the attraction factors
$$
 q(\pi_1) =  0.266 \;,  \qquad q(\pi_2) =  - 0.266 \;  .
$$
However, when pricing, 67\% of subjects found  Lottery $L_2$ more valuable, 
so that
$$
  p(\pi_3) =  0.33 ~ < ~ p(\pi_4) =  0.67 \;  ,
$$
despite that the win in this lottery is less probable. The related 
attraction factors are
$$
q(\pi_3) = - 0.114 \;,  \qquad q(\pi_4) =  0.114 \;   .
$$

Notice that, in the case of pricing, the attraction factor signs are 
reversed as compared to the case of choosing. This is in agreement with the
probability-amount dichotomy (Kim et al., 2012): when choosing, one accepts
as more attractive the lottery with a higher probability win, while when
pricing, one treats as more attractive the lottery with a higher payoff 
amount. In the process of pricing, decision makers usually are more 
pragmatic, evaluating higher the more useful lottery.

Combining the data of this experiment, the two inequalities (\ref{20}) read
$$
 -0.228 < 0.111 < 0.452 \;  ,
$$
which confirms the prediction of QDT.

\vskip 2mm

{\bf Example 2}. When there is no preference reversal, the criterion
(\ref{20}) does not hold. To illustrate this, let us consider an example
treated by Tversky et al. (1990), taking the lotteries
$$
L_1 = \{ 100 , 0.97 \; | \; 0 , 0.03 \} \; , \qquad
L_2 = \{ 400 , 0.31 \; | \; 0 , 0.69 \} \;   .
$$
Their expected utilities are
$$
 U(L_1) = 97 \; , \qquad U(L_2) = 124 \;  ,
$$
which yields the utility factors
$$
 f(\pi_1) = 0.439 \; , \qquad f(\pi_2) = 0.561 \;  .
$$
The first lottery is essentially more certain, and subjects overwhelmingly 
tend to prefer this lottery, so that
$$
 p(\pi_1) =  0.91 ~ > ~ p(\pi_2) =  0.09 \;  .
$$
According to (\ref{4}), the related attractions factors are
$$
 q(\pi_1) =  0.471 \;,  \qquad q(\pi_2) =  - 0.471 \;  .
$$

When pricing, subjects pay higher attention to the payoff amounts so that 
the fraction of decision makers preferring the first lottery is drastically 
reduced. However, the preference reversal does not occur per se, with the 
(more narrow) majority pricing the first lottery higher:
$$
 p(\pi_3) =  0.54 ~ > ~ p(\pi_4) =  0.46 \;  .
$$
The corresponding attraction factors are
$$
 q(\pi_3) =  0.101 \;,  \qquad q(\pi_4) =  - 0.101 \;  .
$$
Since
$$
 f(\pi_2) - f(\pi_1) = 0.122 < 2q(\pi_3) = 0.202 \;  ,
$$
criterion (\ref{20}) is not fulfilled, which is the expected situation
in absence of preference reversal.

This example demonstrates that, although in pricing, one pays a higher
attention to payoff amounts, however the focus is not exclusively on 
this amount. Probabilities can also influence decisions, together with
amounts.

\vskip 2mm

{\bf Example 3}. Another example from Tversky et al. (1990) deals with
the lotteries
$$
L_1 = \{ 12 , 0.92 \; | \; 0 , 0.08 \} \; , \qquad
L_2 = \{ 175 , 0.06 \; | \; 0 , 0.94 \} \;     .
$$
The first lottery is both more certain as well as more useful, with the
expected utilities
$$
U(L_1) = 11.04 \; , \qquad U(L_2) = 10.5
$$
and the utility factors
$$
 f(\pi_1) = 0.513 \; , \qquad f(\pi_2) = 0.487 \;  .
$$
It is not surprising that, when choosing, decision makers prefer this
lottery according to
$$
 p(\pi_1) =  0.81 ~ > ~ p(\pi_2) =  0.19 \;  .
$$
The related attraction factors are
$$
 q(\pi_1) =  0.297 \;,  \qquad q(\pi_2) =  - 0.297 \;  .
$$

When pricing, subjects take into account that the second lottery can 
provide a much higher payoff, yet with too small a probability. As a
result, the fraction of decision makers preferring the first lottery
diminishes, but preference reversal does not happen:
$$
 p(\pi_3) =  0.58 ~  > ~ p(\pi_4) =  0.42 \;   .
$$
In pricing, the first lottery becomes less attractive than in choosing,
but remains more attractive than the second lottery, with the attraction
factors
$$
 q(\pi_3) =  0.067 \;,  \qquad q(\pi_4) =  - 0.067 \;  .
$$
In view of the relations
$$
 f(\pi_2) - f(\pi_1) = -0.026 ~ < ~ 2q(\pi_3) = 0.134 \;   ,
$$
criterion (\ref{20}) does not hold, in agreement with the absence of 
preference reversal. Again, we see that payoff amounts as well as 
probabilities are considered in the process of pricing, although the role 
of payoff amounts, without doubt, is more important in pricing than in 
choosing.

\vskip 2mm

We have also analyzed a large set of data presented by Tversky et al. (1990),
demonstrating the effect of preference reversal. Pairs of lotteries were
presented to 198 participants. In each pair, one of the lotteries, $L_1$,
had a high probability, while the other, $L_2$, a higher payoff with lower
probability. These lotteries are given in Table 1. In each lottery, the
first number is a payoff and the next number is the probability of this
payoff. A lottery is represented as a set $\{x, p(x)\}$, implying that one
gets either the payoff $x$, with probability $p(x)$, or nothing, with
probability $1 - p(x)$. The expected utilities and utility factors are shown.
The first six lottery pairs include rather small payoffs. The following
five pairs contain much larger payoffs by a factor of 25. And the
last five pairs present a mixture of large and small payoffs. All the cases
demonstrate the effect of preference reversal.

In Table 2, we show the prospect probabilities $p(\pi_1)$ and $p(\pi_3)$,
with the corresponding attraction factors $q(\pi_1)$ and $q(\pi_3)$,
demonstrating preference reversal, since $p(\pi_1) > p(\pi_2)$, although
$p(\pi_3) < p(\pi_4)$. Those quantities that are not presented can be found
from the relations
$$
 f(\pi_1) = f(\pi_3) \; , \qquad f(\pi_2) = f(\pi_4) \; ,
$$
$$
 p(\pi_2) = 1 - p(\pi_1) \; , \qquad p(\pi_4) = 1 - p(\pi_3) \; ,
$$
$$
q(\pi_2) = - q(\pi_1) \; , \qquad q(\pi_4) = -q(\pi_3) \; .
$$
We also show the value $[f(\pi_2) - f(\pi_1)]/2$ that has to be compared
with $q(\pi_3)$ and $q(\pi_1)$ in order to check the validity of criterion
(\ref{20}). As is seen from Table 2, the preference reversal criterion
(\ref{20}) is always valid.

Since, in each pair of lotteries considered in the case of choosing or
pricing, the utility factors do not change, the preference reversal effect 
can be interpreted within QDT as caused by the existence of the attraction 
factors. If one would evaluate the lotteries solely on the basis of rational 
utility, no preference reversal would occur. However, preferences of 
decision makers involve irrational feelings and biases as well as other 
considerations not included in the utility, which are embodied in the 
attraction factors, accounting for the phenomenon of preference reversal. 
In order to characterize the deviation from rationality during decision 
making over a family of $N$ trials, we can introduce the 
{\it irrationality measure}
$$
 \dlt_j \equiv \frac{1}{N} \sum_{n=1}^N | q(\pi_j) | \; .
$$
Then $\delta_1$ measures the level of irrationality in the course of 
choosing, while $\delta_3$ describes the degree of irrationality in the 
process of pricing. From Table 2, we find $\delta_1 = 0.299$ and 
$\delta_3 = 0.118$. Thus, people seem to be significantly more irrational 
when choosing, as compared to pricing. In other words, the evaluation of 
lotteries in pricing is more rational.

\section{Discussion}

We have shown that the phenomenon of preference reversal, which is treated
as an anomaly in classical decision making, finds a natural explanation
in the frame of quantum decision theory. In the latter, the preference
probability consists of two terms, the utility factor quantifying the utility
of a prospect, and the attraction factor characterizing behavioral biases of
a decision maker. In that way, a prospect probability, defined as a quantum
quantity, has the meaning of a behavioral probability taking into account
both utility of the considered prospects, as well as their attractiveness
for the decision maker, due to subconscious behavioral biases. We have
formulated the criterion associated within QDT with preference reversal and
we have illustrated its validity for a large set of empirical data.

We summarise the key steps of the logic we have followed.

\begin{enumerate}

\item 
We acknowledge the existence of risk aversion that leads human to prefer 
the more probable outcome ceteris paribus.

\item 
We formulate decisions in terms of QDT and derive the general fundamental 
expression (\ref{4}) of QDT: $p=f+q$.

\item 
We interpret $q$ as an ``attraction factor''  embodying the point resulting 
from risk aversion, which determines the sign of $q$.

\item 
The structure of QDT leads to criterion (20) for preference reversal to 
occur, which relates the utilities of the two lotteries to the attraction 
factors under choosing versus pricing.

\item 
We showed that this criterion is verified by experiments.

\end{enumerate}

We have thus demonstrated that QDT predicts the existence of two 
inequalities for the reversal to occur, that turn out to be confirmed.

It is worth noting that the effect of preference reversal does not only 
occur when choice is compared with pricing, but similar reversals can happen 
in other cases. As another illustration, we can mention the so-called planning 
paradox that can be represented by the following stylized example.

Suppose one is deliberating about stopping smoking. Let the imaginary plan
to stop smoking be denoted as the prospect $\pi_1$, while continuing smoking
corresponds to prospect $\pi_2$. The utility of not smoking clearly overweights 
that of smoking because of evident health reasons. In contrast, the negative 
feelings, connected with addiction, are yet too imaginary to influence the 
mood of the decision maker. We thus expect that the related attraction factors
should be rather small, so that the decision is based mainly on rational grounds.
Hence, the preference in this plan $\pi_1$ is expressed by the inequality
$p(\pi_1) > p(\pi_2)$, implying that the majority of subjects would like to
stop smoking.

However, when one has to choose to really stop smoking now (but not in the 
future), then one actually meets another alternative: really stop smoking, 
which can be denoted as the prospect $\pi_3$, or continue smoking, the prospect 
$\pi_4$. Deciding whether to really stop smoking now, one immediately 
confronts negative feelings anticipating the suffering resulting from addiction. 
This translates into the appearance of a negative attraction factor $q(\pi_3)$ 
devaluating the utility of not smoking. As a result, $p(\pi_3)$ becomes smaller 
than $p(\pi_4)$, which means that the majority of people do not really quit
smoking.

This planning paradox gives a clear example of preference reversal, which
cannot be understood in terms of classical utility considerations, since the 
utility of prospects does not change. But there is no paradox in quantum 
decision theory, where the effect of preference reversal is explained by the
variation of attraction factors. Numerous data, collected by Walsh and
Sanson-Fisher (2001) from the World Health Organization, confirm the robust
existence of the preference reversal in the stop-smoking planning paradox.
Thus the preference reversal is a rather general phenomenon that obtains a
straightforward explanation in the framework of quantum decision theory.

\vskip 2mm

{\bf Conflict of Interests Statement}: There is no conflict of interests.

\vskip 2mm

{\bf Author Contributions}: Both authors equally contributed to the paper.

\newpage

{\Large {\bf References} }

{\parindent=0pt

\vskip 2mm
Ashtiani, M., and Azgomi, M.A. (2015).
A survey of quantum-like approaches to decision making and cognition.
{\it Math. Soc. Sci.} 75, 49--50.

\vskip 2mm
Bagarello, F. (2013).
{\it Quantum Dynamics for Classical Systems}.
Hoboken: Wiley.

\vskip 2mm
Busemeyer, J.R., and Bruza, P. (2012).
{\it Quantum Models of Cognition and Decision}.
Cambridge: Cambridge University.

\vskip 2mm
Busemeyer, J.R., Wang, Z., Khrennikov, A., and Basieva, I. (2014).
Applying quantum principles to psychology.
{\it Phys. Scripta} T163, 014007.

\vskip 2mm
Goldstein, W.M., and Einborn, H.J. (1987).
Expression theory and the preference reversal phenomenon.
{\it Psychol. Rev.} 94, 236--254.

\vskip 2mm
Grether, D.M., and Plott, C.R. (1979).
Economic theory of choice and the preference reversal phenomenon.
{\it Am. Econ. Rev.} 69, 623--638.

\vskip 2mm
Haven, E., and Khrennikov, A. (2013).
{\it Quantum Social Science}. Cambridge: Cambridge university.

\vskip 2mm
Holt, C.A. (1986).
Preference reversal and the independence axiom.
{\it Am. Econ. Rev.} 76, 508--515.

\vskip 2mm
Karni, E., and Safra, Z. (1987).
Preference reversal and the observability of preferences by experimental
methods.
{\it Econometrica} 55, 675--685.

\vskip 2mm
Kim, B.E., Seligman, D., and Kable, J.W. (2012).
Preference reversals in decision making under risk are accompanied by
changes in attention to different attributes.
{\it Front. Neurosci.} 6, 109.

\vskip 2mm
Khrennikov, A. (2010).
{\it Ubiquitous Quantum Structure}.
Berlin: Springer.

\vskip 2mm
Khrennikov, A., and Basieva, I. (2014).
Possibility to agree on disagree from quantum information and decision making.
{\it J. Math. Phsychol.} 62, 1--15.

\vskip 2mm
Lichtenstein, S., and Slovic, P. (1971).
Reversal of preference between bids and choices in gambling decisions.
{\it J. Exp. Psychol.} 89, 46--55.

\vskip 2mm
Lichtenstein, S., and Slovic, P. (1973).
Response-induced reversals of preference in gambling.
{\it J. Exp. Psychol.} 101, 16--20.

\vskip 2mm
Lindman, H.R. (1971).
Inconsistent preferences among gambles.
{\it J. Exp. Psychol.} 89, 390--397.

\vskip 2mm
Loomes, G., and Sugden, R. (1983).
A rationale for preference reversal.
{\it Am. Econ. Rev.} 73, 428--432.

\vskip 2mm
Luce, R.D. (1959).
{\it Individual Choice Behavior: A Theoretical Analysis}.
New York: Wiley. 

\vskip 2mm
Neumann, J. von (1955).
{\it Mathematical Foundations of Quantum Mechanics}.
Princeton: Princeton University.

\vskip 2mm
Schkade, D.A., and Johnson, E.J. (1989).
Cognitive processes in preference reversals.
{\it Org. Behav. Human Perform.} 44, 203--231.

\vskip 2mm
Segal, U. (1988).
Does the preference reversal phenomenon necessarily contradicts the
independence axiom?
{\it Am. Econ. Rev.} 78, 233--236.

\vskip 2mm
Slovic, P., and Lichtenstein, S. (1983).
Preference reversals: a broader perspective.
{\it Am. Econ. Rev.} 73, 596--605.

\vskip 2mm
Sornette, D. (2014).
Physics and financial economics (1776-2014): puzzles,
Ising and agent-based models.
{\it Rep. Prog. Phys.} 77, 062001.

\vskip 2mm
Tversky, A., Sattath, S., and Slovic, P. (1988).
Contingent weighting in judgment and choice.
{\it Psychol. Rev.} 95, 371--384.

\vskip 2mm
Tversky, A., and Thaler, R.H. (1990).
Anomalies: preference reversals.
{\it J. Econ. Persp.} 4, 201--211.

\vskip 2mm
Tversky, A., Slovic, P., and Kahneman, D. (1990).
The causes of preference reversal.
{\it Am. Econ. Rev.} 80, 204--217.

\vskip 2mm
Walsh, R.A., and Sanson-Fisher, R.B. (2001).
{\it Behavioral science learning modules: encouraging people to stop
smoking}. Geneva: World Health Organization.

\vskip 2mm
Yukalov, V.I., and Sornette, D. (2008).
Quantum decision theory as quantum theory of measurement.
{\it Phys. Lett. A} 372, 6867--6871.

\vskip 2mm
Yukalov, V.I., and Sornette, D. (2009a).
Physics of risk and uncertainty in quantum decision making.
{\it Eur. Phys. J. B} 71, 533--548.

\vskip 2mm
Yukalov, V.I., and Sornette, D. (2009b).
Processing information in quantum decision theory.
{\it Entropy} 11, 1073--1120.

\vskip 2mm
Yukalov, V.I., and Sornette, D. (2010).
Mathematical structure of quantum decision theory.
{\it Adv. Complex Syst.} 13, 659--698.

\vskip 2mm
Yukalov, V.I., and Sornette, D. (2011).
Decision theory with prospect interference and entanglement.
{\it Theor. Decis.} 70, 283--328.

\vskip 2mm
Yukalov, V.I., and Sornette, D. (2013).
Quantum probabilities of composite events in quantum measurements with
multimode states.
{\it Laser Phys.} 23, 105502.

\vskip 2mm
Yukalov, V.I., and Sornette, D. (2014a).
Conditions for quantum interference in cognitive sciences.
{\it Top. Cogn. Sci.} 6, 79--90.

\vskip 2mm
Yukalov, V.I., and Sornette, D. (2014b).
Manipulating decision making of typical agents.
{\it IEEE Trans. Syst. Man Cybern. Syst.} 44, 1155--1168.

\vskip 2mm
Yukalov, V.I., and Sornette, D. (2015).
Positive operator-valued measures in quantum decision theory.
{\it Lecture Notes Comput. Sci.} 8951, 146--161.

}

\newpage

{\parindent=0pt
{\bf Table 1}. Pairs of lotteries, with their expected utilities and
utility factors.
}

\vskip 3cm

\begin{center}

{\bf Table 1}

\vskip 5mm

\begin{tabular}{|r|r|c|c|c|c|} \hline
$L_1$    & $L_2$       & $U(L_1)$ & $U(L_2)$ & $f(\pi_1)$ & $f(\pi_2)$\\ \hline
\hline
4,\;0.97   & 16,\; 0.31  &  3.88    & 4.96    & 0.439     & 0.561   \\ \hline
2,\;0.81   &  9,\; 0.19  &  1.62    & 1.71    & 0.486     & 0.514   \\ \hline
3,\;0.94   & 6.5,\; 0.50 &  2.82    & 3.25    & 0.465     & 0.535   \\ \hline
4,\;0.89   & 40,\; 0.11  &  3.56    & 4.4     & 0.447     & 0.553   \\ \hline
2.5,\;0.94 & 8.5,\; 0.39 &  2.35    & 3.315   & 0.415     & 0.585   \\ \hline
2,\;0.92   & 5,\; 0.50   &  1.84    & 2.5     & 0.424     & 0.576   \\ \hline
\hline
50,\; 0.81 & 225,\; 0.19 &  40.5  & 42.75   & 0.486     & 0.514   \\ \hline
75,\; 0.94 & 160,\; 0.50 &  70.5  & 80      & 0.468     & 0.532   \\ \hline
100,\;0.89 & 1000,\;0.11 &  89    & 110     & 0.447     & 0.553   \\ \hline
65,\; 0.94 & 210,\; 0.39 &  61.1  & 81.9    & 0.427     & 0.573   \\ \hline
50,\; 0.92 & 125,\; 0.50 &  46    & 62.5    & 0.424     & 0.576   \\ \hline
\hline
10,\; 0.78 & 100,\; 0.08 &   7.8  & 8       & 0.494     & 0.506   \\ \hline
 7,\; 0.69 &  40,\; 0.17 &  4.83  & 6.8     & 0.415     & 0.585   \\ \hline
 3,\; 0.86 &  13,\; 0.19 &  2.58  & 2.47    & 0.511     & 0.489   \\ \hline
 4,\; 0.94 & 150,\; 0.03 &  3.76  & 4.5     & 0.455     & 0.545   \\ \hline
11,\; 0.89 & 135,\; 0.08 &  9.79  & 10.8    & 0.475     & 0.525   \\ \hline
\hline
\end{tabular}

\end{center}

\newpage

{\parindent=0pt
{\bf Table 2}. Probability $p(\pi_1)$ defined as the fraction
of decision makers choosing the lottery $L_1$, and probability
$p(\pi_3)$ defined as the fraction of subjects pricing the
lottery $L_1$ higher. The corresponding attraction factors $q(\pi_1)$ and
$q(\pi_3)$, and the combination $[f(\pi_2) - f(\pi_1)]/2$ that should be
compared with those attraction factors according to criterion (\ref{20})
obtained from QDT, which reads here  
$q(\pi_3) <  [f(\pi_2) - f(\pi_1)]/2 <  q(\pi_1)$.
}

\vskip 2cm

\begin{center}

{\bf Table 2}

\vskip 5mm

\begin{tabular}{|c|c|c|r|c|} \hline
$p(\pi_1)$ & $p(\pi_3)$ & $q(\pi_1)$ & $q(\pi_3)$ & $[f(\pi_2)-f(\pi_1)]/2$
\\ \hline
\hline
0.83   & 0.26  &  0.391    & $-$0.179    & 0.061   \\ \hline
0.68   & 0.22  &  0.194    & $-$0.266    & 0.014   \\ \hline
0.71   & 0.30  &  0.245    & $-$0.165    & 0.035   \\ \hline
0.71   & 0.33  &  0.263    & $-$0.117    & 0.053   \\ \hline
0.73   & 0.17  &  0.315    & $-$0.245    & 0.085   \\ \hline
0.62   & 0.14  &  0.196    & $-$0.284    & 0.076   \\ \hline
\hline
0.86   & 0.48  &  0.374    & $-$0.006    & 0.014   \\ \hline
0.77   & 0.46  &  0.302    & $-$0.008    & 0.032   \\ \hline
0.84   & 0.47  &  0.393    &  0.023      & 0.053   \\ \hline
0.82   & 0.48  &  0.393    &  0.053      & 0.073   \\ \hline
0.70   & 0.32  &  0.276    & $-$0.104    & 0.076   \\ \hline
\hline
0.81   & 0.38  &  0.316    & $-$0.114    &  0.006   \\ \hline
0.68   & 0.21  &  0.265    & $-$0.205    &  0.085   \\ \hline
0.74   & 0.39  &  0.229    & $-$0.121    & $-$0.011   \\ \hline
0.74   & 0.38  &  0.285    & $-$0.075    &  0.045   \\ \hline
0.79   & 0.46  &  0.315    & $-$0.015    &  0.025   \\ \hline
\hline
\end{tabular}

\end{center}

\end{document}